\documentclass[12pt]{article}

\usepackage{graphicx}
\begin{document}
 
\newpage

\title{\bf{Cosmic Conundrums with Quantum Solutions}}

\author{David E. Rosenberg,\\  EVMS Norfolk VA. 23507 USA \\ }
\date{    }

\maketitle

\section{Abstract}
Dark energy was discovered over 25 years ago and we still do not have an understanding of it. Dark matter comprises about $95\%$ of the matter in the universe and was still do not know what it is made of. The Webb telescope has been finding full formed galaxies with massive black holes in the earliest universe and there is no mechanism for their formation.This paper will explore a quantum density limitation which allows solutions for these and other astrophysical problems. 

Key words: Cosmology theory, dark matter, dark energy, galaxy formation

\section{Understanding the Universe}
Using general relativity, the standard model has matter compressible to a singularity. In gravitational collapses space  must be squeezed out of neutrons and protons. During this process, there is increasing restrictions on particle movement. Particle size becomes a factor. Starting with neutron stars of densities $\approx 10^{14} grams/cm^3$ there has been a loss of gravity (dark energy) between $10-20\%$   \cite{Ponti et al. 2015, Will 2011}. Since each neutron weighs $\approx 1.674927 \times 10^{-24}$ grams, it will take $\approx 10^{38}$ neutrons in each $cm^3$ for neutron star densities. Assuming roughly each neutron is a sphere with radius $10^{-13}$ cm. then each particle is restricted to a space about ten times its size. Protons have similar but slightly smaller sizes. As densities increase over neutron stars, particles begin overlapping and repel each other. The universe's smallest black hole of $4_\odot$ has the highest density of formation $\rho = c^6/G^3M^2$ which equals $10^{17} grams-cm^{-3}$. Here we have about $1000$ more neutrons squeezed into that same $cm^3$ space. Instead of one femtometer (fm.) size, each neutron is squeezed into a space of $0.22$ femtometer. If nucleons could be squeezed further, then further matter packing or densities could occur. However it takes $10^{35}$ pascals to squeeze protons to $0.1$ fm. so even a neutron star can't accomplish this \cite {Burkert et al. 2018}. This is why there are no smaller black holes and the limiting matter density is $\approx 10^{17} grams/cm^3$. Black holes will lose gravity as they collapse and stop collapsing at this level \cite {Bally 2015, Dallilar et al. 2017}.

\section{Explaining Gravitation Loses}
Gravitational masses have been treated as pinpoint sources. However, it is not logical that the gravitational properties of an infinitely collapsing highly squeezed mass would be the same as normal matter.  
For gravitational losses, matter must be highly squeezed and not a gas of noninteracting particles (perfect fluid) nor a Quark-Gluon Plasma. Nucleons highly squeezed to $\approx 0.3$ fm. must be packed so tightly by $10^{17}gm/cm^{3}$ that there is no more remaining space nor motion except quantum jitters at all.
In order to construct a stress-energy tensor \cite{Misner et al. 1973}, the matter 4-velocity $\mathbf{u}=(dt,0,0,0)$, possibly adding only the observer 4-velocity. The 4-momentum is
\begin{equation}
\mathbf{p} = m\mathbf{u}=(m\gamma, m\Delta x \gamma, m\Delta y \gamma, m\Delta z \gamma)  
\end{equation}
where $E=m\gamma$, $\gamma =1$ nonrelativistic and $\Delta x, \Delta y, \Delta z =0$. A volume element $\Sigma$ composed of basis vectors $e_{x},e_{y}, e_{z}$ have zero magnitudes as there is no flow of 4-momentum.   
An observer in his Lorentz frame will measure mass-energy density in $gm/cm^{3} \, T_{00}=T(e_{0},e_{0})$, with the observers 4-velocity $\mathbf{u}$ replaced by $e_{0}$.  
If the box is at rest in the observers frame, all matter kinetic and quantum energy will be sufficiently damped except possibly quantum spin. Space-time interaction will cease, $T_{00}=0$.
To construct a volume in spacetime with a parallelopipid, use four different vectors for edges $\mathbf{A,B,C,D}$. The vectors in standard Lorentz frame are $\mathbf{A}=(\Delta t, 0, 0, 0)$, $\mathbf{B}=(0, \Delta x, 0, 0)$, $\mathbf{C}=(0, 0, \Delta y, 0)$ and $\mathbf{D}=(0, 0, 0, \Delta z)$. 
A 4-volume is
\begin{equation}
\Omega=\epsilon_{\alpha \beta \gamma \delta}A^{\alpha}B^{\beta}C^{\gamma}D^{\delta}=\mathbf{A} \wedge \mathbf{B} \wedge \mathbf{C} \wedge \mathbf{D}
\end{equation}
A volume integral of a tensor \textbf{S} defined over a four dimensional  region $\mathcal{V}$ of spacetime, calculated in a Lorentz frame 
\begin{equation}
M^{\alpha}_{\beta \alpha}=\int S^{\alpha}_{\beta \gamma} dt\, dx \, dy \, dz
\end{equation} 
 The energy density measured in such a volume $E=m\gamma/V=0$  as is the density of the 4-momentum $d\mathbf{p}/dV=0$ per 3 dimensional volume in an observers Lorentz frame.  In the following, $j,k=(1,2, or \,3)$ in what really is a symmetric tensor. $T^{j,0}=0$ is the momentum density, j component. $T^{0,k}=0$ is the energy flux, k component. $T^{j,k}=0$ is the j component of force from matter and fields acting around $x^{k}$.
This keeps the tensor divergence $\mathbf{\nabla} \cdot \mathbf{T}=0$ as there is no particle movement.

With rotating immobile squeezed nucleons, let S be a spacelike hypersurface with arbitrary event $\mathcal{A}$ and coordinates $x^{\alpha}(\mathcal{A})\equiv a^{\alpha}$ using globally inertial coordinates.  
Total angular momentum  on S about $\mathcal{A}$ can be defined as 
\begin{equation}
J^{\mu\nu} \equiv \int_{S} \mathcal{J}^{\mu\nu\alpha} d^{3}\sum_{\alpha}
\end{equation} and will add to total momentum only if present. 
Here 
\begin{equation}
\mathcal{J}^{ \mu\nu\alpha}\equiv(x^{\mu}-a^{\mu})T^{\nu\alpha}-(x^{\nu}-\mathcal{A}^{\nu})T^{\mu\alpha}
\end{equation}
If S is a hyper-surface of constant time t then 
\begin{equation}
J^{\mu\nu}=\int \mathcal{J}^{\mu\nu 0} dx\, dy\, dz
\end{equation}
In the systems rest frame,  let $P^{0}=M$, $P^{j}=0$ and at the center of mass 
\begin{equation}
x_{cm}^{j}=\frac{1}{M}\int x^{j}T^{00}d^{3}x.
\end{equation}
For a large mass, intrinsic angular momentum  is defined angular momentum about any event $(a^{0},x_{cm}^{j})$ on the world line of the center of mass. 
Here components $S^{0j}=0$ and  
\begin{equation}
 S^{jk}=\epsilon^{jkl}S^{l}.\mbox{ and }  S\equiv \int (x-x_{cm})\times d\mathbf{p}/dV \,S^{\mu\nu} d^{3}x
\end{equation} 
The intrinsic angular momentum 4-vector $S^{\mu}$ has components in the rest frame (0,S) 
\begin{equation}
S^{\mu\nu}=U_{\alpha}S_{\beta}\epsilon^{\alpha\beta\mu\nu}
\end{equation} 
The 4-velocity center of a large highly squeezed mass $\mathbf{U}_{\beta}\equiv \mathbf{P}_{\beta}/M =0$.   
Angular momentum is composed of intrinsic and orbital parts. 
An arbitrary event $a$ whose perpendicular distance from the center of mass world line is $-Y^{\alpha}$ making $\mathbf{U}_{\beta}Y^{\beta}=0$.
The total angular momentum $J^{\mu\nu}$ about $\mathcal{A}$ is both the intrinsic part 
\begin{equation}
(S^{\mu\nu}=\mathbf{U}_{\alpha}\mathbf{S}_{\beta}\epsilon^{\alpha\beta\mu\nu})
\end{equation}
 and the orbital part 
\begin{equation}
(L^{\mu\nu}=Y^{\mu} P^{\nu}-Y^{\nu} P^{\mu})
\end{equation}
  With the angular momentum about $\mathcal{A}$ and the 4-momentum known (zero is this case) one can calculate
the vector from $\mathcal{A}$ to the center of mass world line.
\begin{equation}
Y^{\mu}=-J^{\mu\nu}P_{\nu}/M^{2}
\end{equation}
 
Using a swarm of identical particles with event $\mathcal{P}$ inside the swarm, $m_{A}$ is the rest mass. $\mathbf{u}_{A}$ is the 4-velocity, and $\mathbf{p}_{A}$ is the 4-momentum. 
$N_{A}$ is the number of particles per unit volume, as measured in the particles own rest frame. 
The number flux vector 
\begin{equation}
\mathbf{S}_{A}\equiv N_{A}\mathbf{u}_{A}
\end{equation}
 The particles have ordinary velocity $v_{A}$, zero in packed supranuclear densities. 
$\mathbf{u}^{o}_{A}$  is the the Lorentz correction for volume and velocity  $1/(1-v_{A})^{1/2}$. 
The 4-momentum density  is 
\begin{equation}
\mathbf{p}_{A} S^{o}_{A}=m_{A}u^{u}N_{A}u^{o}_{A}
\end{equation} 
Consequently the 4-momentum density has components 
\begin{equation}
T^{uo}_{A}= p^{u}_{A}S^{o}_{A}=m_{A}N_{A}u^{u}_{A}u^{o}_{A}
\end{equation} 
The flux of the $\mu$ component of momentum with perpendicular projection $e_{j}$ is 
\begin{equation}
 T^{\mu j}_{A}= p^{\mu}_{A}S^{j}_{A}=   m_{A}N_{A}u^{\mu}_{A}u^{j}_{A}
\end{equation} 
Here superscripts $(\mu,o)$ and $(\mu,j)$ of the frame independent equation
\begin{equation}
\mathbf{T}_{A}= m_{A}N_{A}\mathbf{u}_{A} \otimes \mathbf{u}_{A}= \mathbf{p}_{A} \otimes \mathbf{S}_{A}
\end{equation}
By summing over all categories, the total number flux vector and stress energy tensor are obtained for all particles in the swarm. If $\mathbf{u}_{A} =0$, these will be zero. 
\begin{equation}
\mathbf{S}=\sum_{A}N_{A}\mathbf{u}_{A} \mbox{ and }\mathbf{T}=\sum_{A} m_{A}N_{A}\mathbf{u}_{A} \otimes\mathbf{u}_{A}=\sum_{A}\mathbf{p}_{A}\otimes\mathbf{S}_{A}
\end{equation} 
The total momentum flux across a closed 3-dimensional surface must vanish $\oint T^{\mu 0} d^{3}\sigma_{a} =0$. 
There is no flux and no sinks and there is no momentum at these supranuclear densities. 

\section{Results of Dark energy and Matter Density Limits}
The quantum correction factor for the stress energy tensor is 0.9 for neutron stars densities and 0.0 at black hole densities $10^{17}grams/cm^3$. Thus it makes no sense to extrapolate general relativity beyond maximum matter density. Gravity is not a property of matter but rather a result of matter movement interacting with space-time. Since the universe never collapsed to a singularity but the limiting density, infinite temperatures and radiation dominance never happened. The observable matter in the universe is $\approx 6 \times 10^{54}$ grams and from nucleosynthesis this is $5\%$ of the total mass. So the universe couldn't collapse $<10^{13}$ scale factor or temperature $>2 \times 10^{13}$ degrees. There was no fireball at the big bang but enough energy to expand the universe. As the core of initial matter was more highly squeezed, it contained most energy. A cold shell and hot core could supply the fully formed galaxies the Webb telescope has found in the early universe. Results pieces of the cold shell coalesced into black holes capturing hot core gases according to the gravitational strength. This would explain the highly correlated galaxies rather than using random process \cite{Cattaneo et al. 2009, Peebles 2011, Rosenberg 2025, Steinhardt 2014, Vandenbergh 2008}. Cold baryons would not have reacted during nucleosynthesis but supplied the gravity to allow the light elements up to lithium to form.

Dark energy is the quantum effects of increasing particle densities at and above $10^{14} grams/cm^3$. Black holes slowly contract to limiting density and stop. The entire pre big bang matter had collapsed to limiting density and stopped. The highly squeezed nucleons powered the re-expansion of the universe. There was no gravity to start. Next is the Friedmann from the Ricci tensor.
\begin{equation}
(a_o/a)^2=-k/a^2+\Omega/3 + 8\pi \rho_m {a_o}^3/3a^3 + 8\pi\rho_r {a_0}^4/3a^4
\end{equation}
Here $a_o$ is the present day universe with the letter m for matter and r for radiation. $\Omega$ is the term Einstein used to keep the universe from expanding and contracting. The k can be $(-1, 0, or +1)$. Because the universe bounced off maximum density, $k=0$. Since the universe matter can't compress to zero, this equation with general relativity must end by maximum density. 

Without dark energy gravitational losses, there could have been no big bang to start the universe. The universe cycles at the big bang. As neutrons were freed, they decomposed to protons and electrons to form the original hydrogen. Radiation had increased to $\approx 2 \times 10^{13}$ degrees but not more. There was a cold shell and hot core to start the universe. The core was more highly squeezed and was an energy sink. Without radiation dominance, initial galaxies formed from the shell pieces coalesced into black holes capturing the following hot core gases. The two stage galaxy formation process preserves angular momentum from the big bang and allows galactic size and speed to be related to the black hole mass.Thus there are similar circular speeds of all galaxies of a given luminosity. The hot and cold matter discrepancies are detectable only below accelerations of $10^{-8}cm-sec^{-2}$ since they are all baryons\cite {Rosenberg 2025}.

There are other examples of gravitation loss over the life of the universe. These are the Hubble tension between  differences in the late and early universe expansion rates and the failure of galactic clumping over expected early universe structure. Both are due to black hole collapses to maximum density with result gravitation losses. Supernova bouncing off maximum density cores no longer require solely neutrinos to explode.

\section{Discussion}
General relativity has been extrapolated in black holes and the big bang to 
enormous energies and densities without consideration of properties of matter. Lightly squeezing nucleons like neutron stars will result in a light reduction of gravitational energy. Highly squeezing nucleons as in pre-big bang matter and black holes will result in elimination of gravitational energy.  
Highly squeezing matter will cause quantum effects in the stress-energy tensor so that $T_{\mu\nu}\rightarrow \mathbf{0}$. Very simple 'quantum gravity' can be produced. The gravitational losses including dark energy can't be explained any other way. A cold shell and hot core produced
the initial Planck spectrum radiation. The cold shell dispersion led to the great wall, filaments and voids.
The basis of almost all galaxies formed simultaneously as cold dark shell matter coalesced into black holes. which captured subsequent hot core gases into proto-galaxies. The formed stars ionized the intergalactic medium. The fact that the Higgs Boson is $125$ GeV and not larger, eliminates non baryons as dark matter. 

Gravitation is described by the Riemann metric.  No negative energies are required nor is a vacuum energy necessary for empty space.
The fact that initial universe had flat space-time and a Planck spectrum has been explained. 
Since the origin of the big bang is a bounce, it should re-collapse into the big crunch. The resulting neutrons will eventually decompose into the protons and electrons to make hydrogen and helium for the next cycle.   
Black holes must continuously accrete mass-energy to maintain 
their gravitational strength. Gravity is not caused by matter itself but rather by the motion of matter particles.  Galaxies were made in a two stage process.  First the big bang shell laid down all halos and super-massive black holes. Then hot core gases were captured according to the size of each gravitational well.

\twocolumn

\endthebibliography{55}

\end{document}